\documentstyle[aps]{revtex}
\begin{document}
\newcommand{\beq}{\begin{equation}}
\newcommand{\eeq}{\end{equation}}
\newcommand{\beqn}{\begin{eqnarray}}
\newcommand{\eeqn}{\end{eqnarray}}
\newcommand{\bmath}{\begin{mathletters}}
\newcommand{\emath}{\end{mathletters}}
\twocolumn[\hsize\textwidth\columnwidth\hsize\csname @twocolumnfalse\endcsname
\title{Comment on ``Discovery of microscopic electronic inhomogeneity
in the high-$T_c$ superconductor $Bi_2Sr_2CaCu_2O_{8+x}$'', cond-mat/0107347}
\author{J. E. Hirsch }
\address{Department of Physics, University of California, San Diego\\
La Jolla, CA 92093-0319}
 
\date{\today} 
\maketitle 
\begin{abstract} 
The paper by Pan et al, cond-mat/0107347 \cite{pan}, reports the discovery
of 'electronic inhomogeneity' and 'spatial variations in the local density
of states' in  high $T_c$ cuprates. In my paper ``Effect of local potential
variations in the model of hole superconductivity'', Physica C {\bf 194}, 119
(1992)\cite{local}, many of the findings of Pan et al were foreshadowed.
\end{abstract}
\vskip2pc]

 The model of hole superconductivity\cite{hole} was proposed in 1989 as an
explanation of high $T_c$ superconductivity in cuprates. The abstract
of ref. 2, ``Effect of local potential
variations in the model of hole superconductivity'' reads: ``In the model of hole 
superconductivity the strength of the pairing interaction depends on the
local carrier density. This gives rise to a dependence of the gap
function $\Delta_k$ on the band energy $\epsilon_k$. Fluctuations in the
local potential energy will result in different values of $\Delta_k$ at
the Fermi energy and hence in different values of the local energy gap.
In particular, the energy gap can be sharply reduced. We study this
behavior by numerical solution of the Bogoliubov-de Gennes equations for
the model. The behavior is contrasted with what occurs in the attractive
Hubbard model, where local potential fluctuations have negligible effect.
The physical origin of this behavior and the possible relevance to high
$T_c$ oxides is discussed.''

Pan et al\cite{pan} report the discovery of 'spatial variations in both the 
local density of states (LDOS) spectrum and the superconducting energy gap'. They
find that these variations occur 'on a surprisingly short
lengt scale of $14 \AA$', which 'appears shorter than
the experimental in-plane superconducting coherence length
$\xi_{ab}\sim 22-27 \AA$'. They propose that their observations
'naturally lead one to relate the magnitude of the integrated
local density of states to the local oxygen doping concentration',
and they find that 'the spectra obtained at points with larger integrated
LDOS values exhibit higher differential conductance, smaller gap
values, and sharper coherence peaks'.

In summary, Pan's et al results\cite{pan} suggest that in the regions
of higher oxygen doping concentrations the hole concentration in the plane
is larger and this leads to the observed variations, on a surprisingly
short length scale.

In our 1992 paper\cite{local}, we explored one of the consequences of the 
finite slope of the gap function predicted in the theory of hole 
superconductivity\cite{hole}. The paper says
'we study the effect of local potential fluctuations in the
local energy gap (as measured in a tunneling experiment), by numerical
solution of the Bogoliubov-de Gennes equations for
the model. We find that such fluctuations can cause large variations in the
local energy gap'. In that paper we modeled the potential fluctuations by variations in
the local single particle energy $\epsilon_i$, and considered both
situations where '$\epsilon_i$ varies slowly over distances of the
order of the coherence length', and situations where
'$\epsilon_i$ varies over distances of the order of the coherence length
or shorter'. We found that 'similar sensitivity of the local energy
gap to disorder is expected in both regimes of slowly varying and
rapidly varying fluctuations with respect to the coherence length'.
We plotted the behavior of the local tunneling density of states,
arguing that 'with a scanning tunneling microscope and a tip of atomic 
dimensions it is in principle possible to obtain atomic resolution'.
We concluded 'our results illustrate that tunneling measurements 
at different points in a sample can exhibit large variations due to variations
in the local potential energy'. We found the surprising result that 'for disorder on length
scales of the order of or shorter than the coherence length, variations
in the energy gap will occur due to the finite slope of the 
gap function'. 

Finally, we gave physical arguments to explain these
effects: 'A change in on-site energies changes the strength of the
interaction in the model of hole superconductivity due to the fact that for higher
(lower) on-site energies the wavefunction at the Fermi energy becomes more
bonding (antibonding) like'. We contrasted this with the behavior of other
models, in particular the attractive Hubbard model, were no such behavior is expected
and concluded 'we would expect real materials described by the model of hole
superconductivity to be more sensitive to nonmagnetic disorder than those
described by attractive Hubbard or similar models, particularly if the
slope of the gap function is large'. We also discussed the relation of this to
the kinetic energy gain predicted to occur in this model upon going superconducting.

Figure 1(a) shows one of the figures of our '92 paper (a)\cite{local}, and 
Figure 1(b) show spectra
reported by Pan et al\cite{pan}. The reader will note an asymmetry
in the spectra of opposite sign in both panels. This is because in our
paper $\omega$ refers to the voltage of the tip,
while in Pan's paper the bias voltage refers to the voltage of the 
sample. The theory of hole superconductivity predicts larger 
conductance for negatively biased sample\cite{hole}. The different curves
in Figure 1(a) correspond to different values of the on-site
energy $\epsilon_i$ for the hole: in going from the solid to the
dash-dotted curve, $\epsilon_i$ becomes more negative, corresponding
to an $increase$ in the local hole concentration; this will occur in
the presence of higher oxygen doping in the region. The reader will
note that the gap becomes smaller and the coherence peaks become
sharper, a similar behavior as seen in the progression from curves
1 to 5 in (b), the data of Pan et al. According to Pan et al\cite{pan},
those curves correspond to increasing LDOS values, corresponding to
higher local oxygen doping concentration.

The spectra shown in Figs. 1(a) and 1(b) do not agree in every detail.
Nevertheless, we believe the qualitative similarity, and the fact that
the results in (a) were published 9 years ago, are significant.
It is also interesting to note that the paper of Pan et al\cite{pan} makes no
reference to the work discussed here\cite{local}, even though both the first and
last authors of Ref. 1  were made aware of that work. The model of hole
superconductivity makes many other predictions on the superconductivity
of high $T_c$ cuprates, $MgB_2$, and other materials\cite{last}.

\begin{figure}
\caption { (a) Numerical results for the local density of states
in the presence of local potential variations in the model of
hole superconductivity, from Ref. 2. In the
progression from solid to dash-dotted curves the local hole site
energy $\epsilon_i$ becomes more negative, leading to increased
local hole concentration. (b) Experimental data from Pan et al, ref. 1.
The progression from curves 1 to 5 corresponds to increasing LDOS, 
corresponding to increasing local hole concentration.}
\label{Fig. 1}
\end{figure}


\begin{references}
\bibitem{pan} S.H. Pan et al, cond-mat/0107347 (2001).
\bibitem{local} J.E Hirsch, Physica C {\bf 194}, 119 (1992).
\bibitem{hole} J.E. Hirsch and F. Marsiglio, Phys. Rev. B {\bf 39}, 11515
(1989);  Physica C {\bf 162-164}, 591 (1989); F. Marsiglio and J.E. Hirsch,
Phys. Rev. B {\bf 41}, 6435 (1990); J. E. Hirsch, 
Physica C {\bf 158}, 326 (1989).
\bibitem{last} J.E. Hirsch, cond-mat/0106310 (2001).


\end{references}
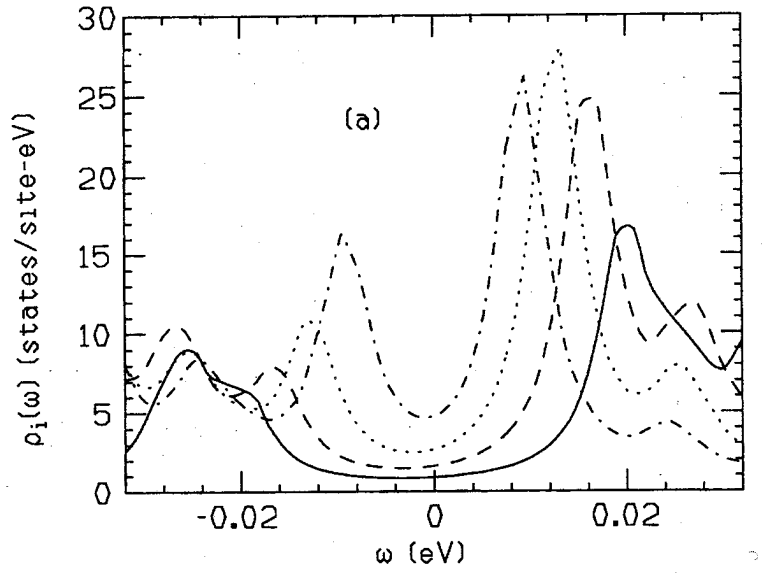
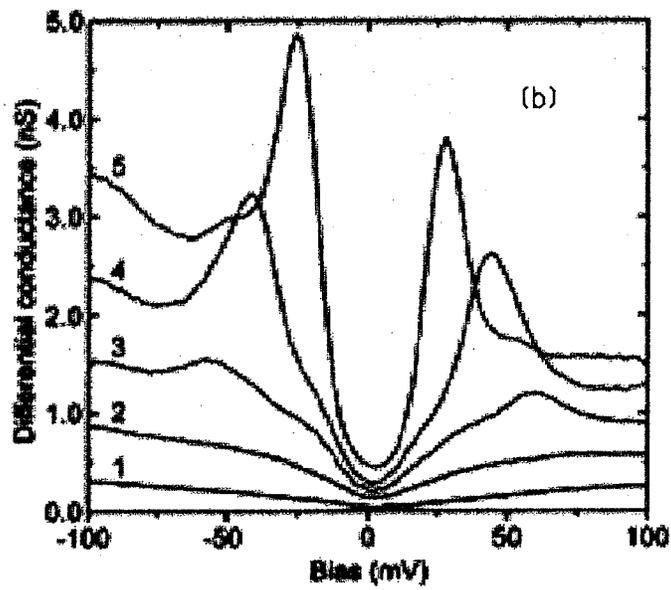
\end{document}